\documentclass{rrparticle}
\usepackage{graphicx}
\usepackage{color}

\def\be{\begin{equation}}
\def\ee{\end{equation}}
\def\bea{\begin{eqnarray}}
\def\eea{\end{eqnarray}}

\title{Vortex solutions in atomic Bose-Einstein condensates via the Adomian
Decomposition Method
}
\author[1a,1b,1c]{Tiberiu Harko}
\author[2]{Man Kwong Mak}
\author[3]{Chun Sing Leung}
\affil[1a]{Astronomical Observatory, 19 Ciresilor Street,  Cluj-Napoca 400487, Romania,  Email$^a$: {\em tiberiu.harko@aira.astro.ro}}
\affil[1b]{Faculty of Physics, Babes-Bolyai University, Kogalniceanu Street,
Cluj-Napoca 400084, Romania,}
\affil[1c]{School of Physics, Sun Yat-Sen University, Guangzhou 510275, People's
Republic of China}
\affil[2]{Departamento de F\'{\i}sica, Facultad de Ciencias Naturales, Universidad de
Atacama, Copayapu 485, Copiap\'o, Chile, Email: {\em mankwongmak@gmail.com}}
\affil[3]{Department of Applied Mathematics,
Hong Kong Polytechnic University, Hong Kong, Hong Kong SAR, P. R. China, Email: {\em  chun-sing-hkpu.leung@polyu.edu.hk }}


\begin{document}
\maketitle
\begin{abstract}
We study the dynamics of vortices with arbitrary topological charges in weakly interacting Bose-Einstein condensates using the Adomian Decomposition Method to solve the nonlinear Gross-Pitaevskii equation in polar coordinates. The solutions of the vortex equation are expressed in the form of infinite power series.  The power series representations are compared with the exact numerical solutions of the Gross-Pitaevskii equation for the uniform and the harmonic potential, respectively. We find that there is a good agreement between the analytical and the numerical results.
\end{abstract}

\maketitle

\tableofcontents

\section{Introduction}

An important theoretical and physical phenomenon, the Bose-Einstein condensation, has received considerable attention especially after its experimental realization in trapped atomic gases \cite{Pethick, Gr, new1,new2,new3}.
The basic
principle in the understanding of the Bose-Einstein Condensates (BECs) is that at
very low temperatures all integer spin particles occupy the lowest quantum
state of the system. Hence in this temperature range microscopic quantum
phenomena dominates the dynamics. One of the most interesting properties of Bose-Einstein condensates is the way they behave under rotation.  At higher rotational frequencies the angular momentum generates vortex filaments at which the superfluid density vanishes, while
the circulation of the velocity field around a closed contour which encircles the vortex is quantized \cite{Pethick, Gr,new1,new2,new3}. The first experimental observation of vortices in BECs \cite{vore} has also led to intensive theoretical researches in this field  \cite{RMP}.

From a theoretical point of view BECs confined by an external potential are described by the Gross-Pitaevskii equation (GPE) \cite{Pethick, Gr}.  Similarly to other nonlinear dispersive equations, the GPE supports various types of solitary wave solutions \cite{RMP}. In particular, in the two-dimensional setting the GPE has vortex solutions of the form $\Psi (r,t)=e^{-i\mu t}e^{il\theta}\psi (r)$, where $(r,\theta)$ are polar coordinates, $l$ is the vortex degree, $\mu $ is the chemical potential, and $\psi(r)$ is the non-negative radial vortex profile \cite{RMP}. Since the GPE is a strongly non-linear differential equations, it is generally not possible to obtain its solutions in an exact analytical form.

It is the goal of the present paper to consider the vortex solutions of the two-dimensional GPE by using the Adomian Decomposition Method (ADM) \cite{R1,R2,b2}, which allows to obtain semianalytical solutions of many types of ordinary,
partial and integral differential nonlinear equations.  The ADM usually generates the solution of a differential equation in the form of
a series, with the terms of the series determined recursively by using the
Adomian polynomials.  Reviews of the ADM in applied
mathematics and on its applications in physics can be found in \cite{R1,R2},
respectively. The ADM  has been used in a wide range of scientific investigations
\cite{C5,C6,C7,C8,C9,C10,C11, C12}.

In order to apply the ADM for the study of vortices we transform first the GPE in polar coordinates into an equivalent integral equation. Then, by decomposing the nonlinear term in the GPE into a series of polynomials of the form $\sum _{n=1}^{\infty}{A_n}$, where $A_n$ are the so-called
Adomian polynomials, the power series solution of the GPE can be obtained in a general representation, and for arbitrary external potentials. We apply our analytic results to the case of the uniform and harmonic potentials, and in each case the semianalytic solution is compared with the exact numerical solution.

The present paper is organized as follows. We introduce the basic concepts of the Bose-Einstein Condensation and of vortex dynamics in Section~\ref{sect1}. The ADM is briefly reviewed in Section~\ref{sect2}, where the power series solution of the GPE in polar coordinates is also presented. The ADM solutions for the uniform and harmonic potentials are obtained in Section~\ref{sect3}, and they are compared with the exact numerical results. We discuss and conclude our results in Section~\ref{sect4}.

\section{Bose-Einstein Condensation and vortex dynamics}\label{sect1}

In the present Section we briefly review the basic concepts on the Bose-Einstein Condensation, and the dynamics of the vortices.

\subsection{The Gross-Pitaevskii equation}

 The Bose-Einstein Condensation process is essentially
determined by the quantum mechanical correlation between the particles in
the gas. At high temperatures the de Broglie thermal wavelength is greater
than the mean interparticle distance. When the temperature $T$ of the boson
gas becomes lower than the critical one, $T\leq T_{cr}$, the transition to
the condensate phase begins. The critical temperature $T_c$ is given by $T_{cr}=2\pi\hbar^2\rho_{cr}^{2/3}/ \zeta^{2/3}\left(3/2\right)m^{5/3}k_{B}$ \cite{Pethick,Gr},
where $\rho_{cr}$ is the critical transition density, $m$ is the mass of the
particle forming the condensate, $k_{B}$ is Boltzmann's constant, and $\zeta $
denotes the Riemmann zeta function, respectively.

 Systems of bosons confined in a nonuniform potential can be investigated by using the GPE, which can be derived from the variational principle $\delta \left(H-\mu N_0\right)/\delta \Psi^{*}=0$ \cite{Pethick,Gr},
 where $N_0$ denotes the total particle number, and $H$ is the GP
energy functional of the system, given by
\begin{eqnarray}
H\left[\Psi,\Psi^{*}\right]= \int{\left[\frac{\hbar ^2}{2m}\left|\nabla
\Psi\right|^2+V_{ext}\left|\Psi\right|^2+\frac{g}{2}\left|\Psi\right|^4%
\right]d\vec{r}},
\end{eqnarray}
where $g=4\pi \hbar ^2a/m$, and $a$ is the scattering length. The variation gives the
GPE as
\begin{equation}  \label{GP}
\left[-\frac{\hbar ^2}{2m}\Delta +V_{ext}\left(\vec{r}\right)+g\left|\Psi%
\right|^2\right]\Psi \left(\vec{r}\right)=\mu \Psi\left(\vec{r}\right).
\end{equation}

Eq. (\ref{GP}) is a nonlinear Schr\"{o}dinger equation, whose cubic term
accounts for the contact interaction between bosons. The Gross-Pitaevskii
wave function is normalized according to $\int{\left|\Psi\left(\vec{r}%
\right)\right|^2d\vec{r}}=N_0$. 


\subsection{Vortex dynamics}

Due to the experimental advances in the study of
atomic Bose-Einstein condensates, the investigations of the
dynamics of quantized vortices, few-vortex clusters and large scale vortex
lattices have seen a tremendous development \cite{v1,v2,v3,v4}. A vortex is a
topological property of a superfluid. If one takes a closed path around the
vortex, the phase of the wave function undergoes a $2\pi$ winding, and,
consequently, the flow of the superfluid is quantized. One of the simplest
cases in which formation of vortices may occur is the two-dimensional case
with circular symmetry \cite{Pethick}. Then in the coordinate system $\left(r, \theta,t\right)$ the GPE can be
represented as $\Psi\left(r,\theta, t\right)=\mathcal{R}(r){\rm Exp}\left(i l \theta\right){\rm Exp}\left(-i \mu t/\hbar\right)$,
with the radial part $\mathcal{R}(r)$ satisfying the equation \cite{Pethick}
\be\label{f}
-\frac{\hbar ^2}{2m}\left[\frac{d^2\mathcal{R}(r)}{dr^2}+\frac{1}{r}%
\frac{d\mathcal{R}(r)}{dr}\right]+\frac{\hbar ^2l^2}{2mr^2}\mathcal{R}(r)+
V_{ext}(r)\mathcal{R}(r)+g\mathcal{R}^3(r)=\mu \mathcal{R}(r).
\ee

The solutions of Eq. (\ref{f}) that behave near the origin $r\rightarrow 0$
as $R(r)\sim r^l$ are the vortex solutions \cite{Pethick}. The existence and properties of
such solutions have been intensively investigated in the physical
literature. For example, if $V_{ext}(r)=0$, for $r\rightarrow \infty $, $R(r)$
tends to a constant, $R(r)\rightarrow \sqrt{n_0}$ \cite{Fe}. On the other
hand, in the case of the harmonic potential $V_{ext}(r)=m\omega _0^2r^2/2$, the
solutions of Eq.~(\ref{f}) behave like $R(r)\sim
P\left(r/a_0\right){\rm Exp}\left(-r^2/2a_0^2\right)$, where $P\left(r/a_0\right)$ is a
polynomial in $r/a_0$, and the oscillator length $a_0$ is given by $%
a_0=\left(\hbar/m\omega _0\right)^{\frac{1}{2}}$ \cite{Sols}. The complex order
parameter $\left|\Psi \left(\vec{r}\right)\right|$ must be a single valued
function, implying that the index $l$ in Eq.~(\ref{f}) must be an integer
number, which is called the topological charge. The integer nature of $l$
follows directly from the quantization of circulation, $\oint {\vec{v}\cdot d%
\vec{s}}=\oint{\frac{\hbar}{2m}\nabla \phi \cdot d\vec{s}}=\frac{\hbar }{m}l$%
, where $\phi \equiv \mathrm{arg}\left(\Psi\right)$ \cite{Pethick}. The quantization of a
circulation is the basic property of quantum vortices.

In the following we rescale the
independent and dependent variables according to $r=\sqrt{\hbar ^{2}/2m\mu }x$ and $\mathcal{R}%
=\sqrt{\mu /g}R$, respectively. By denoting $v=V_{ext}/\mu $, Eq.~(\ref{f}) takes the
form
\begin{equation}
\frac{d^{2}R(x)}{dx^{2}}+\frac{1}{x}\frac{dR(x)}{dx}-\left[ \frac{l^{2}}{x^{2}}%
+\left( v(x)-1\right) \right] R(x)-R^{3}(x)=0.  \label{f1}
\end{equation}%
We will consider Eq.~(\ref{f1}) together with the boundary conditions $%
R(0)=R_{0}=0$, and $R\left(x_b\right)=R_b$, respectively.

\section{Power series solutions for vortex lines via the
Adomian Decomposition Method}\label{sect2}

In the following we present the basic formalism of the ADM, and we apply it to obtain power series solutions of the vortex equation.

\subsection{The Adomian Decomposition Method}

The Adomian decomposition method can be summarized as follows. Consider a
differential equation of the form
\begin{equation}
Lu(x)+Nu(x)=g(x),
\end{equation}%
where $L$ is a linear operator, $N$ represents the non-linear terms, and $%
g\left( x\right) $ is a source term. Applying the inverse operator to both
sides of the above equation we obtain
\begin{equation}
u(x)=L^{- 1}g(x)-L^{-1}\left[ Nu(x)\right] .
\end{equation}

The function $u(x)$ is decomposed as $u(x)=\sum_{n=1}^{\infty }u_{n}(x)$,
while the non- linear operator $Nu(x)$ is decomposed by an infinite series
of Adomian polynomials, corresponding to the specific non linearity, so that
$Nu(x)=\sum_{k=1}^{\infty }A_{k}(x)$,
where the Adomian polynomials $A_n$ are constructed based on the following
algorithm,
\begin{equation}
A_{k}(x)=\frac{1}{k!}\frac{d^{k}}{d\varepsilon ^{k}}\left[ N\left(
\sum_{n=0}^{\infty }\varepsilon ^{n}u_{n}\right) \right] |_{\varepsilon =0}.
\end{equation}



Let's consider now the general second order equation
\begin{equation}
f^{\prime \prime }(x)+a(x)f^{\prime }(x)+b(x)f(x)=h(x),  \label{f3}
\end{equation}%
and let $\varphi \left( x\right) \neq 0$ be a solution of the homogeneous
part of the equation. Then, by using the method of the variation of the
parameters, the general solution of Eq. (\ref{f3}) is given by \cite{Grama}
\begin{eqnarray}\label{f3a}
f(x)=C_{1}\varphi (x)+C_{2}\int \frac{dx}{E(x)\varphi
^{2}(x)}+
\varphi \left( x\right) \int \frac{dx}{E(x)\varphi ^{2}(x)}%
\int E(x)\varphi (x)h(x)dx,
\end{eqnarray}%
where $E(x)=\exp \left[ \int a(x)dx\right] $, and $C_{1}$ and $C_{2}$ are
arbitrary constants of integration.

\subsection{Power series  solutions of the vortex equation}

In order to obtain a power series solution of  Eq. (\ref{f1}) 
we write the equation in the form
\begin{equation}
\frac{d^{2}R(x)}{dx^{2}}+\frac{1}{x}\frac{dR(x)}{dx}- \frac{l^{2}}{x^{2}} R(x)=\left[v(x)-1\right]R(x)+R^3(x).  \label{f4}
\end{equation}

The left hand (homogeneous) side of the above equation,
\begin{equation}
\frac{d^{2}R_{0}(x)}{dx^{2}}+\frac{1}{x}\frac{dR_{0}}{dx}- \frac{l^{2}%
}{x^{2}} R_{0}(x)=0,
\end{equation}
has the general solution $R_{0}^{\pm l}(x)=c_{\pm}x^{\pm l}$, where $c_{\pm}$ are arbitrary integration constants.
In order to avoid any singular
behavior at the center we adopt as the particular solution of the equation $%
R_{0}(x)=c_{+}x^l$.  By taking into account that $E(x)=x$, and by taking $C_2=0$,  Eq. (\ref%
{f4}) can be reformulated as an equivalent integral equation given by
\begin{eqnarray}
R(x)=x^l\Bigg\{c_++\int_0^x{x^{-1-2l}dx}\int_0^x{x^{1+l}\left[\left(v(x)-1\right)R(x)+R^3(x)\right]dx}\Bigg\}.
\end{eqnarray}

We decompose now $R(x)$ as $R(x)=\sum_{n=0}^{\infty }R_{n}(x)$ and $R^{3}=\sum_{n=0}^{\infty }A_{n}(x)$, where $A_n(x)$ are the Adomian polynomials. Hence we obtain the following recursive relations for the successive determination of the terms $%
R_{n}(x)$,
\begin{eqnarray}
&&R_{0}(x)=c_+x^l, \;R_{k+1}(x)=-x^{l}\int_{0}^{x}x^{-1-2l}dx\int_{0}^{x}x^{1+l}R_{k}(x)dx+
\notag \\
&&x^{l}\int_{0}^{x}x^{-1-2l}dx\int_{0}^{x}x^{1+l}\left[ v(x)R_{k}(x)+A_{k}(x)%
\right] dx,\;
k =0,1,2,...
\end{eqnarray}

The solution of the equation is given by
$R\left( x\right) =R_{0}+R_{1}+R_{2}+R_{3}+...$, and it can be obtained as
\bea
R\left( x\right) &=&c_+x ^{l}-\sum_{k=0}^{\infty
}x^{l}\int_{0}^{x}x^{-1-2l}dx\int_{0}^{x}x^{1+l}R_{k}(x)dx+\nonumber\\
&&\sum_{k=0}^{%
\infty }x^{l}\int_{0}^{x}x^{-1-2l}dx\int_{0}^{x}x^{1+l}\left[
v(x)R_{k}(x)+A_{k}(x)\right] dx.
\eea

The first few Adomian polynomials for the function $R^{3}$ are
\begin{equation}
A_{0}(x)=R_{0}^{3}(x),
A_{1}(x)=3R_{1}(x)R_{0}^{2}(x),
\end{equation}%
\begin{equation}
A_{2}(x)=3R_{2}(x)R_{0}^{2}(x)+3R_{1}^{2}(x)R_{0}(x),
\end{equation}%
\begin{equation}
A_{3}(x)=3R_{3}(x)R_{0}^{2}(x)+6R_{1}(x)R_{2}(x)R_{0}(x)+R_{1}^{3}(x).
\end{equation}

\section{Applications: the cases of the uniform and of the harmonic potentials}\label{sect3}

In the present Section we will consider the power series solutions of the vortex equation for the case of a uniform potential, which for simplicity we will take as zero, and for the case of the harmonic potential, with $V_{ext}(r)=m\omega _0^2r^2/2$.

\subsection{The uniform potential $v(x)=0$}

In the case of the vanishing exterior potential $v(x)=0$, and  the vortex equation (\ref{f1}) takes the form
\begin{equation}
\frac{d^{2}R(x)}{dx^{2}}+\frac{1}{x}\frac{dR(x)}{dx}- \frac{l^{2}}{x^{2}} R(x)=-R(x)+R^{3}(x).
\end{equation}
The first few terms in its series solution  can be obtained as
\be
R_0(x)=c_+x^l, R_1(x)=\frac{c_+^3 x^{3 l+2}}{4 (l+1) (2 l+1)}-\frac{c_+ x^{l+2}}{4 (l+1)},
\ee
\bea
R_2(x)&=&\frac{3 c_+^5 x^{5 l+4}}{32 (l+1)^2 (2 l+1) (3 l+2)}-\frac{c_+^3 (3 l+2) x^{3 l+4}}{16 (l+1)^2 (l+2) (2 l+1)}+\nonumber\\
&&\frac{c_+ x^{l+4}}{32
   (l+1) (l+2)},
\eea
\bea
R_3(x)&=&\frac{c_+^7 (12 l+7) x^{7 l+6}}{128 (l+1)^3 (2 l+1)^2 (3 l+2) (4 l+3)}-\nonumber\\
&&\frac{c_+^5 \left(30 l^2+57 l+26\right) x^{5 l+6}}{128 (l+1)^3 (l+2)
   (2 l+1) (2 l+3) (3 l+2)}-\frac{c_+
   x^{l+6}}{384 (l+1) (l+2) (l+3)}+\nonumber\\
 &&  \frac{c_+^3 \left(18 l^2+45 l+19\right) x^{3 l+6}}{128 (l+1)^2 (l+2) (l+3) (2 l+1) (2 l+3)}.
\eea

\begin{figure}[h]
 \centering
\includegraphics[scale=0.65]{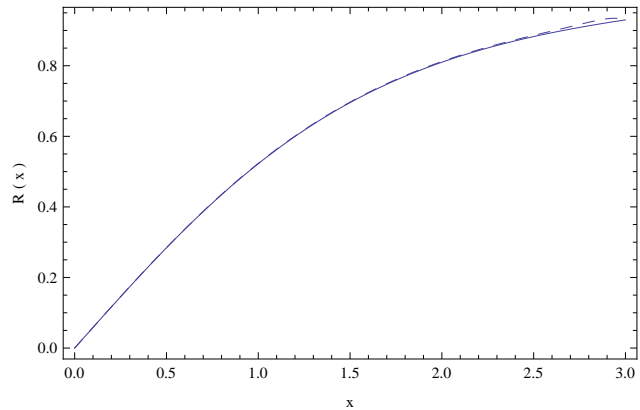}
\caption{Comparison of the exact numerical solution of the vortex equation (\ref{f1}) (solid curve) and of the Adomian truncated power series solution (dashed curve) with $k=8$ for the case of the uniform potential $v(x)=0$, and for $l=1$. The boundary conditions used to solve the equation are $R(0)=0$, and $R(3)=0.93$. }\label{fig1}
\end{figure}

The comparison between the exact numerical solution of the vortex equation and the Adomian truncated power series representation with $k=8$ is presented, for $l=1$, in Fig.~\ref{fig1}. The adopted boundary conditions are $R(0)=0$, and $R(3)=0.93$, respectively. With the use of the boundary conditions the numerical value of the integration constant $c_+$ is obtained as $c_+=0.58690$.

As one can see from the Figure, the Adomian power series solution gives a good description of the numerical results.

\subsection{The harmonic potential $v(x)=x^2$}

As a second case of the comparison between the ADM and the numerical approaches we consider the case of the harmonic potential $V_{ext}(r)=m\omega _0^2r^2/2$. After rescaling the variable $r$, the harmonic potential becomes $V_{ext}=\left(\omega _0^2\hbar ^2/4\mu\right)x^2$. By fixing the chemical potential so that $\mu =\hbar \omega _0/2$, we obtain $v(x)=V_{ext}/\mu=x^2$. For the harmonic potential the vortex equation becomes
\begin{equation}
\frac{d^{2}R(x)}{dx^{2}}+\frac{1}{x}\frac{dR(x)}{dx}- \frac{l^{2}}{x^{2}} R(x)=\left(x^2-1\right)R(x)+R^3(x).
\end{equation}

A few terms in the Adomian series expansion of the solution of the vortex equation for the harmonic potential for $l=1$ can be obtained as
\be
R_0(x)=c_+x, R_1(x)=\frac{1}{24} c_+ \left(c_+^2+1\right) x^5-\frac{c_+ x^3}{8},
\ee
\bea
R_2(x)=-\frac{1}{576} c_+ \left(5 c_+^2+2\right) x^7+\frac{c_+ \left(3 c_+^4+4 c_+^2+1\right) x^9}{1920}+\frac{c_+ x^5}{192},
\eea
\bea
R_3(x)&=&\frac{c_+ \left(41 c_+^2+5\right) x^9}{46080}+\frac{c_+ \left(c_+^2+1\right) \left(19 c_+^4+16 c_+^2+1\right)
   x^{13}}{322560}-\nonumber\\
 &&  \frac{c_+ \left(339 c_+^4+302 c_+^2+23\right) x^{11}}{691200}-\frac{c_+ x^7}{9216},
\eea
\bea
R_4(x)&=&-\frac{c_+ \left(163 c_+^2+5\right) x^{11}}{2764800}+\frac{c_+ \left(4467 c_+^4+2596 c_+^2+49\right)
   x^{13}}{58060800}+\nonumber\\
 &&  \frac{c_+ \left(c_+^2+1\right) \left(619 c_+^6+845 c_+^4+253 c_+^2+3\right)
   x^{17}}{278691840}-\nonumber\\
  && \frac{c_+ \left(6681 c_+^6+9450 c_+^4+3023 c_+^2+44\right) x^{15}}{270950400}+\frac{c_+ x^9}{737280},
\eea
\bea
R_5(x)&=&\frac{c_+ \left(5197 c_+^2+35\right) x^{13}}{1857945600}-\frac{c_+ \left(103881 c_+^4+37834 c_+^2+154\right)
   x^{15}}{13005619200}+\nonumber\\
 &&  \frac{c_+ \left(606495 c_+^6+650622 c_+^4+131266 c_+^2+409\right) x^{17}}{117050572800}+\nonumber\\
&& \frac{c_+
   \left(c_+^2+1\right) \left(51351 c_+^8+97040 c_+^6+53218 c_+^4+7304 c_+^2+15\right) x^{21}}{613122048000}-\nonumber\\
 &&  \frac{c_+
   \left(4081205 c_+^8+7913244 c_+^6+4512834 c_+^4+667364 c_+^2+1689\right) x^{19}}{3511517184000}-\nonumber\\
 &&  \frac{c_+
   x^{11}}{88473600}.
\eea

\begin{figure}[h]
 \centering
\includegraphics[scale=0.65]{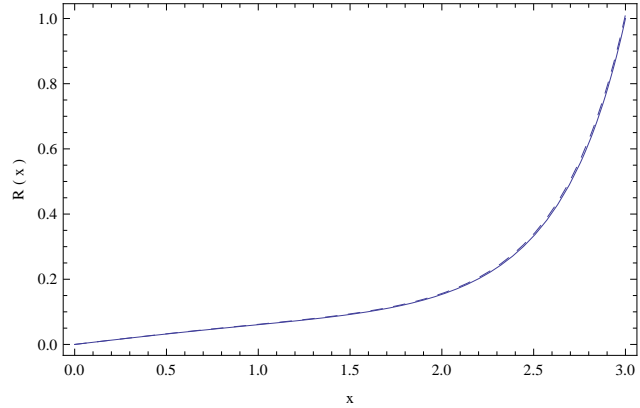}
\caption{Comparison of the exact numerical solution of the vortex equation (\ref{f1}) (solid curve) and of the Adomian truncated power series solution (dashed curve) with $k=8$ for the case of the harmonic potential $v(x)=0$, and for $l=1$. The boundary conditions used to solve the equation are $R(0)=0$, and $R(3)=0.93$. }\label{fig2}
\end{figure}

The comparison between the exact numerical solution and the ADM power series solution, truncated to nine terms, is presented in Fig.~\ref{fig2}. For the boundary conditions we have adopted the numerical values $R(0)=0$, and $R(3)=0.93$, which fix the integration constant $c_+$ as $c_+=0.06711$. As one can see from Fig.~\ref{fig2}, there is a good agreement between the numerical solution and the analytical one.

\section{Discussions and concluding remarks}\label{sect4}

Vortices represents the breakdown of the laminar fluid flow, and thus they also  play a fundamental role in turbulence. The vortex is associated with the fluid rotation, which can be described by the circulation  $\Gamma =\oint {\vec{v}\left(\vec{r}\right)\cdot d\vec{r}}$
around the vortex, where $\vec{v}\left(\vec{r}\right)$
is the velocity field of the fluid. Classical vortices can have any numerical value of the circulation. On the other hand  superfluids are irrotational, and any rotation or angular momentum occurs through vortices described by a quantized circulation. A key role in the dissipation of transport in superfluids is played by quantized vortices. Many forms of quantized vortices in BECs have been observed experimentally, like, for example, single vortices, vortex pairs and rings, or vortex lattices.

In the present paper we have considered some power series solutions of the Gross-Pitaevskii equation describing the behavior of quantized vortices, by using the Adomian Decomposition Method, representing a powerful mathematical method for solving nonlinear differential and functional
equations. The ADM is essentially based on the decomposition of the
solution of the nonlinear operator equation into a series of functions. In order to apply the ADM we have first reformulated the GPE as a Volterra type integral equation. Then the solution of the Volterra equation can be obtained straightforwardly in the form of an infinite power series. We have considered two cases of physical interest, corresponding to the choice of the external potential as a uniform one, and as the harmonic potential, respectively. For each case the analytic results were compared with the exact numerical solution, and it was found that there is an excellent agreement between the analytic ADM solution and the numerical one. The present approach could lead to new insights into the important problem of the vortex structure, since other important physical quantities, like, for example, the particle number and the energy per unit length can be easily obtained in an analytical form.

\section*{Acknowledgments}

We would like to thank to the anonymous referee for comments and suggestions that helped us to improve our manuscript.

\end{document}